# Energy-dependent kinetic model of photon absorption by superconducting tunnel junctions

G. Brammertz
*Science Payload and Advanced Concepts Office, ESA/ESTEC, Postbus 299, 2200AG Noordwijk, The Netherlands.*

A. G. Kozorezov, J. K. Wigmore
*Department of Physics, Lancaster University, Lancaster, United Kingdom.*

R. den Hartog, P. Verhoeve, D. Martin, A. Peacock
*Science Payload and Advanced Concepts Office, ESA/ESTEC, Postbus 299, 2200AG Noordwijk, The Netherlands.*

A. A. Golubov, H. Rogalla
*Department of Applied Physics, University of Twente, P.O. Box 217, 7500 AE Enschede, The Netherlands.*

**Abstract** – We describe a model for photon absorption by superconducting tunnel junctions in which the full energy dependence of all the quasiparticle dynamic processes is included. The model supersedes the well-known Rothwarf-Taylor approach, which becomes inadequate for a description of the small gap structures that are currently being developed for improved detector resolution and responsivity. In these junctions relaxation of excited quasiparticles is intrinsically slow so that the energy distribution remains very broad throughout the whole detection process. By solving the energy-dependent kinetic equations describing the distributions, we are able to model the temporal and spectral evolution of the distribution of quasiparticles initially generated in the photo-absorption process. Good agreement is obtained between the theory and experiment.

PACS numbers: 85.25.Oj, 74.50.+r, 74.40.+k, 74.25.Jb

# I. Introduction

The development of superconducting tunnel junctions (STJs) for application as photon detectors for astronomy and materials analysis continues to show great promise[1,2]. Over the last few years both the quality of the devices fabricated and the understanding of the relevant physical processes have improved greatly. Notable amongst the latter are the proximity effect which determines the properties of the bi-layer electrodes commonly used[3,4,5], the various diffusion and loss mechanisms which limit the energy resolution of an STJ[6,7] and the details of the quasiparticle interactions through which the overall performance of the detector is modelled[8,9,10]. In this paper we describe a major advance in the treatment of quasiparticle dynamics, which is essential for modeling the latest generation of low gap, multi-tunneling STJs designed to operate at mK temperatures.

Previously the response of a biased STJ to the absorption of a photon creating non-equilibrium quasiparticles has commonly been modelled within the framework of the Rothwarf-Taylor balance equations[11]. The main assumption of this model is that during the initial down-conversion process quasiparticles relax very rapidly to the superconducting edge. Further stages of charge transfer, loss and recombination are evaluated under the assumption that all active quasiparticles reside at the superconducting edge and hence that all have this same energy. However, even in experiments involving large gap STJs based on Nb or Ta, evidence was found that the mean energy of the quasiparticles lay above the superconducting edge, and that the energy distribution remained relatively broad during the whole current integration time[12,13]. Lower gap multi-tunneling STJs, using for example Al, are now being developed to take advantage of the intrinsically higher resolution and responsivity. In such devices the relaxation times of excited quasiparticles are greatly increased to the point where it is impossible to describe the experimental results with an over-simplified mono-energetic model.

In this paper we give a description of STJ photon detection that includes the full energy dependence of tunneling, relaxation and loss processes. The model is presented for the most general type of STJ, one in which the two electrodes are not BCS-type superconductors but are proximised. Most commonly devices are of the form S/Al/AlOx/Al/S, where the higher gap superconductor S takes the role of photon absorber and the Al is primarily to facilitate the growth of a high quality insulating barrier between the electrodes. Such electrodes have properties intermediate between those of bulk Al and bulk S, and cannot be accurately described by the simpler BCS relationships. However the BCS forms can easily be retrieved from the expressions given. We begin from the kinetic equations for the quasiparticle numbers as a function of energy in both electrodes, and solve them to obtain the quasiparticle energy distributions as a function of time. From these we go on to determine the various parameters of interest and compare the results with experiments on a proximised Nb/Al device.

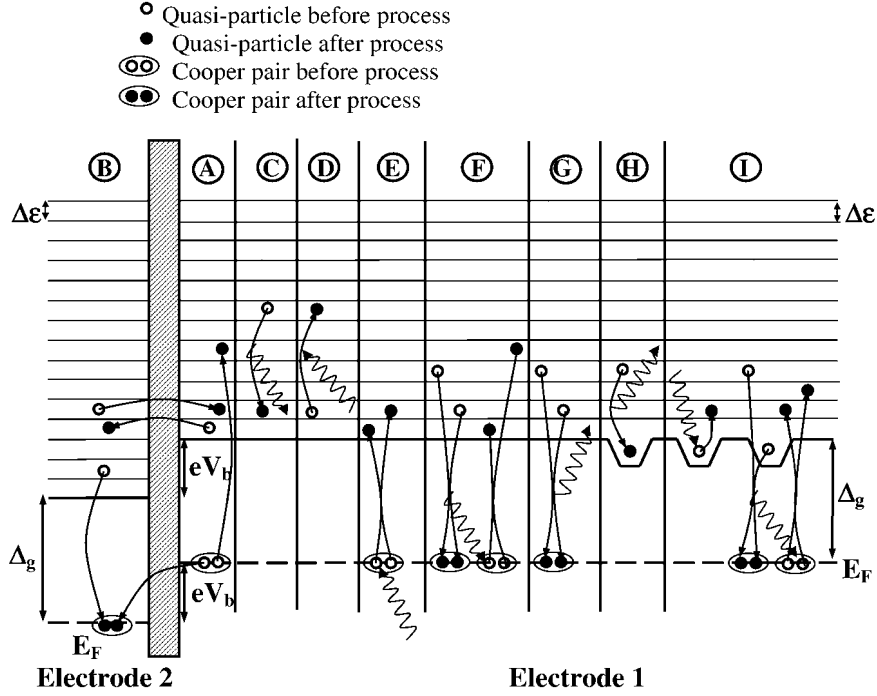

**Figure 1:**
Schematical semiconductor representation of all processes included in our kinetic equations model. **A:** Tunnelling and back-tunnelling. The direct tunnel event is the one going from electrode one to electrode two. The back-tunnelling event is the more complicated process involving Coopers pairs. **B:** Cancellation tunnelling. The cancellation tunnel event is the one going from electrode two to electrode one. The cancellation back-tunnelling is not shown for simplicity. It can be found by reversing the arrows on the back-tunnelling schematic of A. **C:** Electron-phonon scattering with emission of a phonon (quasiparticle relaxation). **D:** Electron-phonon scattering with absorption of a phonon (quasiparticle excitation). **E:** Cooper pair breaking. **F:** Quasiparticle recombination with energy exchange. **G:** Quasiparticle recombination with phonon loss. **H:** Quasiparticle trapping by relaxation. **I:** Quasiparticle de-trapping by phonon absorption and by recombination with an untrapped quasiparticle.

## II. Characteristic rates

In order to determine the quasiparticle energy distribution we have to solve the energy-, position- and time-dependent kinetic equations[14] for the quasiparticle numbers in both electrodes. We are interested in modeling the evolution of the quasiparticle distribution starting from the moment when the generated quasiparticles fill homogeneously the whole volume of the electrode and hence we neglect the lateral gradients in the kinetic equations. Nevertheless, the position dependence comes from the fact that the detector is not homogeneous in the direction perpendicular to the barrier. The time it takes for a quasiparticle to traverse the width of the electrode (~1psec) is much faster than any of the quasiparticle processes occurring in the junction. For this reason we can average the kinetic equations over the position perpendicular to the barrier. This removes the position dependence in our final expression of the energy dependent kinetic equations.

Many processes occur in the electrodes of our detectors, for which characteristic rates must be calculated. Figure 1 shows a semiconductor representation of all processes included in our model.

In order to be able to calculate the characteristic rates of all the processes shown in fig. 1, we need to know several basic characteristics of the bi-layer forming the electrodes of the detector. These characteristics are: the normalized energy and position dependent density of states in the electrode i (i = 1,2) $DoS_i(x,\varepsilon)$, the imaginary part of the Green's function of the bi-layer $ImF_i(x,\varepsilon)$, and the position dependent order parameter $\Delta_i(x)$. All three quantities can be calculated with the proximity effect theory[3,4,15]. Here, x is the direction perpendicular to the barrier and $\varepsilon$ is the quasiparticle energy. Figure 2 shows the real (density of states) and the imaginary part of the Green's function of a Nb-Al bi-layer with 100 nm of Nb and 120 nm of Al at 4 different positions in the bi-layer. The variation of the order parameter as a function of position in the bi-layer is shown as well. At low enough temperatures (typically $T<T_C/10$) all three quantities are independent of the temperature of the superconductor. We recall that the main results of the proximity effect theory applied to bi-layers with film thickness of the order of the coherence length in the material are a uniform energy gap $\Delta_g$ across the bi-layer and a strong discontinuity of all quantities at the interface between the two materials. More details can be found in Refs. [3,4,15].

Of course, our model is also valid for homogeneous junctions, for which the electrodes obey BCS relationships. In this case the results of the proximity effect theory have to be replaced by their BCS counterparts in all the equations:

$$DoS(x,\varepsilon) \rightarrow \frac{\varepsilon}{\sqrt{\varepsilon^2 - \Delta_g^2}} \quad (1)$$

$$ImF(x,\varepsilon) \rightarrow \frac{\Delta_g}{\sqrt{\Delta_g^2 - \varepsilon^2}} \quad (2)$$

$$\Delta(x) \rightarrow \Delta_g \quad (3)$$

In the BCS case the position dependence disappears.

Using the results of the proximity effect theory we can then calculate the energy dependent characteristic rates of all processes occurring in biased STJs (Fig. 1). Since the kinetic equation for quasiparticles in superconductors is non-linear due to the presence of recombination collision integrals, it cannot in general be analyzed analytically for the case of significant deviations from the equilibrium state. Therefore, we will use a numerical approach. For this purpose we divide the energy domain into $N_{en}$ energy intervals of an arbitrary width $\delta\varepsilon$, which can be made as small as one prefers, with the cost of increased calculation time. We choose a range of the energy domain from $\Delta_g$ to $4\Delta_g$ typically divided into 30 intervals, a choice which is usually a good compromise between acceptable calculation time and sufficient accuracy.

In the following we will explicitly calculate the characteristic rates of the different processes. All rates will be calculated for electrode 1, but are of course also valid for electrode 2 by interchanging the indices 1 and 2.

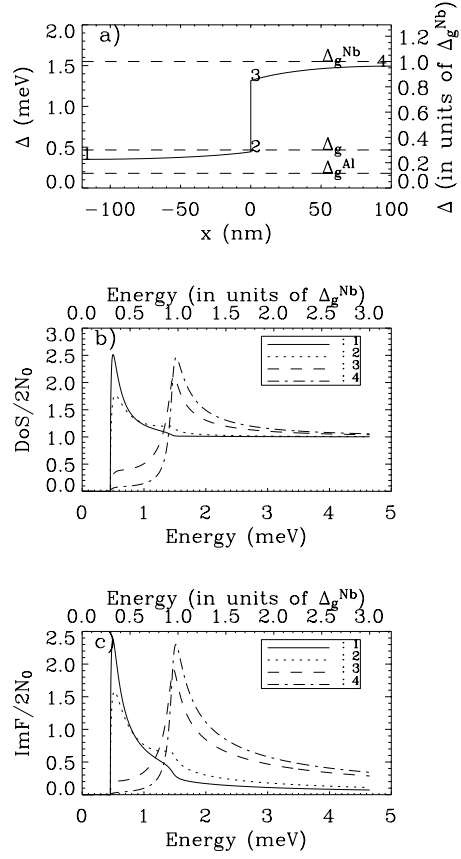

**Figure 2:**
(a) Pair potential Δ for a Nb-Al bi-layer with 100 nm of Nb and 120 nm of Al. The upper dashed line is the bulk energy gap of Nb. The lower dashed line is the bulk energy gap of Al. The intermediate dashed line is the energy gap of the bi-layer, as determined from the density of states. The points 1, 2, 3 and 4 correspond to the four positions in the bi-layer for which the density of states is given in (b). **(b)** Density of states DoS for a Nb-Al bi-layer with 100 nm of Nb and 120 nm of Al. The densities of states are represented for both materials at the free interfaces and at the Nb-Al interface. The points 1 to 4 in (a) indicate the positions in the bi-layer for which the densities of states are given. **(c)** Imaginary part of the Green's function ImF for a Nb-Al bi-layer with 100 nm of Nb and 120 nm of Al. The imaginary part of the Green's function ImF is given at the same four positions in the bi-layer as for the density of states.

### A. Forward tunneling

The forward (i. e. with energy gain), tunneling rate from an energy $\varepsilon_\alpha$ in electrode 1 to an energy $\varepsilon_\alpha + eV_b$ in electrode 2 is given by[12,15,16,17,18]:

$$\Gamma_{tun}(\varepsilon_\alpha \rightarrow \varepsilon_\alpha + eV_b) = \frac{1}{4eR_nA} \frac{DoS_1(0,\varepsilon_\alpha)DoS_2(0,\varepsilon_\alpha + eV_b)}{\int_{electr1} N_0(x)DoS_1(x,\varepsilon_\alpha)dx}, \qquad (4)$$

where $N_0$ is the single spin density of states at the Fermi energy in the normal state (Its dependence on x is to indicate that the material is not the same throughout the electrode), $R_n$ is the normal resistance of the junction, A is the area of the junction and $V_b$ is the positive potential difference between electrodes 1 and 2. Values of all material parameters appearing in (4)-(20) are summarized in table I for Nb and Al.

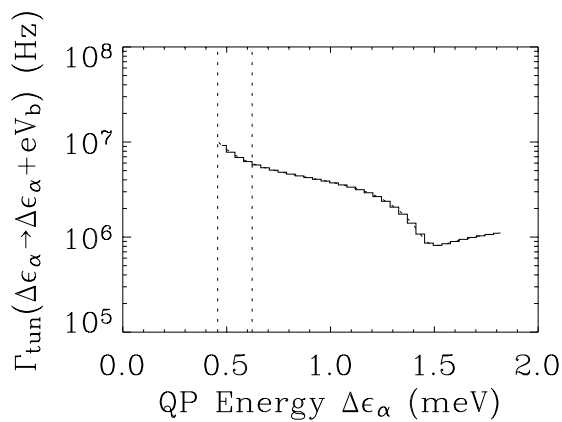

**Figure 3:**
Direct tunnel rate as a function of quasiparticle energy. The solid line gives the tunnel rate as a function of energy intervals $\Gamma_{tun}(\Delta\varepsilon_\alpha)$, whereas the dashed line shows the tunnel rate as a function of quasiparticle energy $\Gamma_{tun}(\varepsilon_\alpha)$. Note that the two lines are pretty much superimposed. The first vertical dotted line indicates the energy gap, whereas the second vertical dotted line indicates the bias energy above the gap.

**Table I. Parameters used for Nb and Al.**

| Symbol | Name | Unit | Nb | Al |
|---|---|---|---|---|
| $R_nA$ | Normal resistivity of junction | $\mu\Omega\ cm^2$ | 2.35 $\pm$ 0.2 | |
| $T_C$ | Critical temperature | K | 9.4 | 1.2 |
| $\Delta_g$ | Energy gap | $\mu eV$ | 1550 | 180 |
| $N_0$ | Single spin normal state density of states at Fermi energy | $10^{27}$ states $eV^{-1}\ m^{-3}$ | 31.7 | 12.2 |
| $\alpha^2$ | Average square of the electron-phonon interaction matrix element | meV | 4.6 | 1.92 |
| N | Ion number density | $10^{28}\ m^{-3}$ | 5.57 | 6.032 |
| $\tau_0$ | Electron-phonon interaction characteristic time | nsec | 0.149 | 440 |
| T | Temperature | K | 0.3 | |

The notation ($\varepsilon_\alpha \to \varepsilon_\beta$) indicates that during the process the quasiparticle changes its energy from $\varepsilon_\alpha$ at the beginning of the process to $\varepsilon_\beta$ at the end of the process. This notation will be the same throughout the paper.

We can now determine the mean forward tunneling rate in the energy interval $\Delta\varepsilon_\alpha$, where $\Delta\varepsilon_\alpha$ is the interval $[\varepsilon_\alpha-\delta\varepsilon/2, \varepsilon_\alpha+\delta\varepsilon/2]$, by integrating $\Gamma_{tun}(\varepsilon_\alpha \to \varepsilon_\alpha+eV_b)$ over $\varepsilon$ in $\Delta\varepsilon_\alpha$ and dividing by $\delta\varepsilon$:

$$\Gamma_{tun}(\Delta\varepsilon_\alpha \to \Delta\varepsilon_\alpha + eV_b) = \frac{\int_{\Delta\varepsilon_\alpha} \Gamma_{tun}(\varepsilon \to \varepsilon + eV_b) d\varepsilon}{\delta\varepsilon} \qquad . \qquad (5)$$

Figure 3 shows the tunnel rate for a 100 nm Nb and 120 nm of Al junction as function of energy intervals for a bias voltage of 180 µV. The tunneling rate is independent of temperature, under the condition that the density of states is independent of temperature (typically $T<T_C/10$). One can see that a non-essential technical limitation of the model is that it only works for bias energies $eV_b$ which are an integer multiple of the energy interval $\Delta\varepsilon$, because otherwise the quasiparticle ends up in between energy intervals.

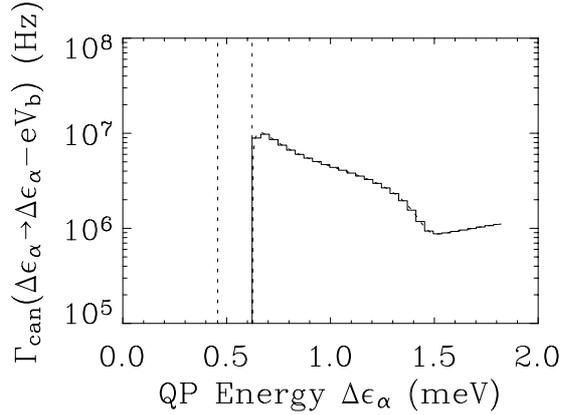

**Figure 4:**
Cancellation tunnel rate as a function of quasiparticle energy. The solid line gives the tunnel rate as a function of energy intervals $\Gamma_{can}(\Delta\varepsilon_\alpha)$, whereas the dashed line shows the tunnel rate as a function of quasiparticle energy $\Gamma_{can}(\varepsilon_\alpha)$. Note that the two lines are pretty much superimposed. The first vertical dotted line indicates the energy gap, whereas the second vertical dotted line indicates the bias energy above the gap.

## B. Cancellation tunneling

When the quasiparticles have an energy $eV_b$ above the gap energy, they can tunnel against the direction indicated by the bias voltage. During these tunnel processes the quasiparticles will lose an energy $eV_b$ and will create a current in the direction opposite to the forward tunnel currents. For these cancellation tunnel events the following equation holds[12,15,16,17,18]:

$$\Gamma_{can}(\varepsilon_\alpha \to \varepsilon_\alpha - eV_b) = \frac{1}{4eR_nA} \frac{DoS_1(0,\varepsilon_\alpha)DoS_2(0,\varepsilon_\alpha - eV_b)}{\int_{electr1} N_0(x)DoS_1(x,\varepsilon_\alpha)dx}, \quad (6)$$

Note that the cancellation rate is zero for quasiparticle energies lower than $\Delta_g+eV_b$, simply because no states are available at the corresponding energies in electrode 2.

The cancellation rate in the energy interval $\Delta\varepsilon_\alpha$ can be written similarly to (5).

$$\Gamma_{can}(\Delta\varepsilon_\alpha \to \Delta\varepsilon_\alpha - eV_b) = \frac{\int_{\Delta\varepsilon_\alpha} \Gamma_{can}(\varepsilon \to \varepsilon - eV_b)d\varepsilon}{\delta\varepsilon}. \quad (7)$$

Figure 4 shows the cancellation tunnel rate for a 100 nm Nb and 120 nm of Al junction as a function of energy intervals for a bias voltage of 180 µV.

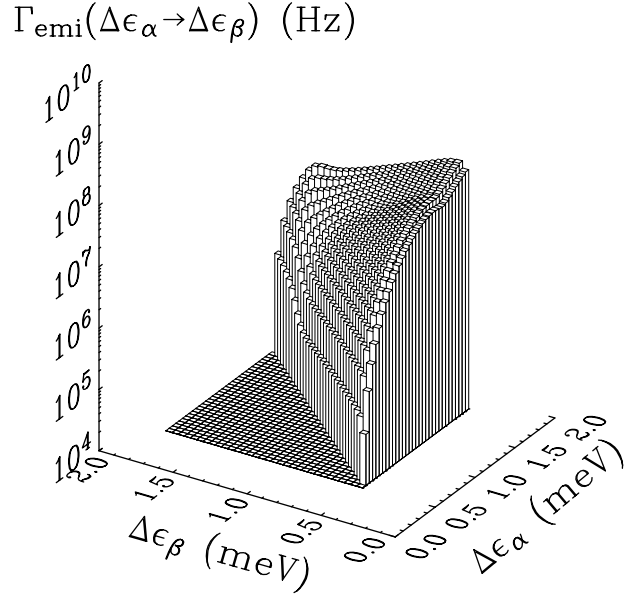

**Figure 5:**
Electron-phonon scattering with emission of a phonon rate, $\Gamma_{emi}(\Delta\varepsilon_\alpha \to \Delta\varepsilon_\beta)$, as a function of initial and final quasiparticle energy. Note that above the diagonal all rates are zero, because the quasiparticle cannot go to a higher energy after phonon emission. The phonon emission rate is proportional to the emitted phonon energy cubed.

### C. Rate for electron-phonon scattering with emission of a phonon (Relaxation)

A quasiparticle can scatter from energy $\varepsilon_\alpha$ to a lower energy $\varepsilon_\beta$ by emitting a phonon of energy $\varepsilon_\alpha - \varepsilon_\beta$. The mean rate for the relaxation of a quasiparticle of energy $\varepsilon_\alpha$ at the position x in the electrode to the energy interval $\Delta\varepsilon_\beta$, by emission of a phonon of energy $\varepsilon_\alpha - \varepsilon_\beta$ is given by[16,19]:

$$\Gamma_{emi}(x,\varepsilon_\alpha \to \Delta\varepsilon_\beta) = \frac{1}{\tau_0(x)[kT_C(x)]^3} \int_{\varepsilon_\alpha-\varepsilon_\beta-\delta\varepsilon/2}^{\varepsilon_\alpha-\varepsilon_\beta+\delta\varepsilon/2} \Omega^2 \left[ DoS_1(x,\varepsilon_\alpha - \Omega) - \frac{\Delta_1(x)}{\varepsilon_\alpha} Im F_1(x,\varepsilon_\alpha - \Omega) \right] [1+n(\Omega)] d\Omega$$
(8)

where $\tau_0$ is a material constant defined in ref. [19], $T_C$ is the bulk critical temperature of the material and $n(\Omega)$ is the phonon distribution function.

This phonon emission rate still depends on the position in the bi-layer x. As already stated previously, transport over the vertical direction is much faster than the typical time constant of phonon emission. Therefore, one can average the phonon emission rate over the vertical position in the bi-layer x:

$$\Gamma_{emi}(\varepsilon_\alpha \to \Delta\varepsilon_\beta) = \frac{\int_{electr1} N_0(x) DoS_1(x,\varepsilon_\alpha) \Gamma_{emi}(x,\varepsilon_\alpha \to \Delta\varepsilon_\beta) dx}{\int_{electr1} N_0(x) DoS_1(x,\varepsilon_\alpha) dx}.$$
(9)

Finally, we can average the phonon emission rate from an energy $\varepsilon_\alpha$ into the energy interval $\Delta\varepsilon_\beta$ over the energy interval $[\varepsilon_\alpha-\delta\varepsilon/2, \varepsilon_\alpha+\delta\varepsilon/2]$, in the same way as was done for the tunnel and the cancellation rate.

$$\Gamma_{emi}(\Delta\varepsilon_\alpha \to \Delta\varepsilon_\beta) = \frac{\int_{\Delta\varepsilon_\alpha} \Gamma_{emi}(\varepsilon_\alpha \to \Delta\varepsilon_\beta) d\varepsilon_\alpha}{\delta\varepsilon}.$$
(10)

Figure 5 shows the phonon emission rate as a function of the initial energy of the quasiparticle $\Delta\varepsilon_\alpha$ and as a function of the quasiparticle energy after phonon emission $\Delta\varepsilon_\beta$ for the Nb-Al bilayer.

### D. Electron-phonon scattering with absorption of a phonon (Excitation)

A quasiparticle can scatter from an energy $\varepsilon_\alpha$ to a higher energy $\varepsilon_\beta$ by absorbing a phonon of energy $\varepsilon_\beta-\varepsilon_\alpha$. The mean rate for the excitation of a quasiparticle of energy $\varepsilon_\alpha$ at the position x in the electrode to the energy interval $\Delta\varepsilon_\beta$, by absorption of a phonon of energy $\varepsilon_\beta-\varepsilon_\alpha$ is given by[16,19]:

$$\Gamma_{abs}(x,\varepsilon_\alpha \to \Delta\varepsilon_\beta) = \frac{1}{\tau_0(x)[kT_C(x)]^3} \int_{\varepsilon_\beta-\varepsilon_\alpha-\delta\varepsilon/2}^{\varepsilon_\beta-\varepsilon_\alpha+\delta\varepsilon/2} \Omega^2 \left[ DoS_1(x,\varepsilon_\alpha + \Omega) - \frac{\Delta_1(x)}{\varepsilon_\alpha} Im F_1(x,\varepsilon_\alpha + \Omega) \right] n(\Omega) d\Omega$$
(11)

The next two steps are the same as for the previous section. The position independent phonon absorption rate is determined by:

$$\Gamma_{abs}(\varepsilon_\alpha \to \Delta\varepsilon_\beta) = \frac{\int\limits_{electr1} N_0(x) DoS_1(x,\varepsilon_\alpha) \Gamma_{abs}(x,\varepsilon_\alpha \to \Delta\varepsilon_\beta) dx}{\int\limits_{electr1} N_0(x) DoS_1(x,\varepsilon_\alpha) dx}. \qquad (12)$$

And finally we average $\Gamma_{abs}(\varepsilon_\alpha \to \Delta\varepsilon_\beta)$ over the energy interval $[\varepsilon_\alpha-\delta\varepsilon/2, \varepsilon_\alpha+\delta\varepsilon/2]$:

$$\Gamma_{abs}(\Delta\varepsilon_\alpha \to \Delta\varepsilon_\beta) = \frac{\int\limits_{\varepsilon_\alpha-\delta\varepsilon/2}^{\varepsilon_\alpha+\delta\varepsilon/2} \Gamma_{abs}(\varepsilon_\alpha \to \Delta\varepsilon_\beta) d\varepsilon_\alpha}{\delta\varepsilon} \qquad (13)$$

For the situation when the phonon distribution is not disturbed and remains in equilibrium $n(\Omega)$ is the Planckian distribution and the described rate is the scattering rate with absorption of a thermal phonon. The phonon absorption rate in this case is strongly temperature dependent. The excess phonons created by the photon absorption and quasiparticle relaxation in the biased STJ are not considered.

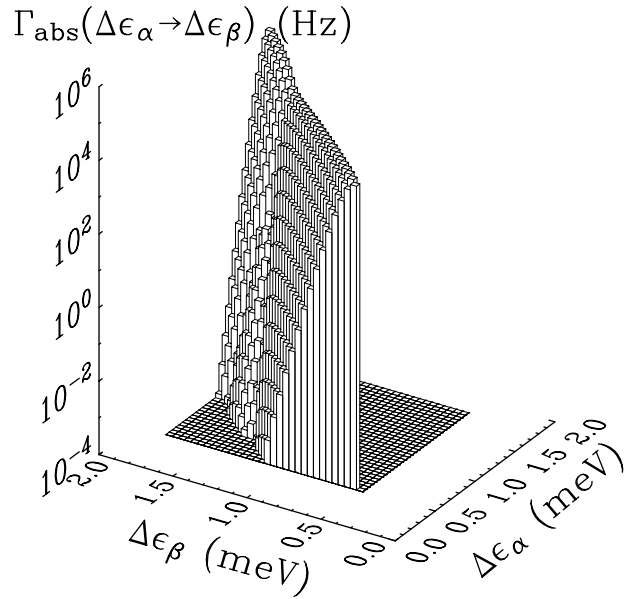

**Figure 6:**
Electron-phonon scattering with absorption of a phonon rate, $\Gamma_{abs}(\Delta\varepsilon_\alpha \to \Delta\varepsilon_\beta)$, as a function of initial and final quasiparticle energy. Note that below the diagonal all rates are zero, because the quasiparticle cannot go to a lower energy after phonon absorption. The phonon absorption rate decreases exponentially with increasing phonon energy, because of the exponential dependency of the Bose distribution function.

Figure 6 shows the thermal phonon absorption rate as a function of the initial energy of the quasiparticle $\Delta\varepsilon_\alpha$ and as a function of the quasiparticle energy after phonon absorption $\Delta\varepsilon_\beta$ for the Nb-Al bilayer at a temperature of 300mK.

### E. Cooper pair breaking

The rate at which a phonon of energy $\Omega > 2\Delta_g$ breaks a Cooper pair into two quasiparticles is given by[12,19]:

$$\Gamma_{PB}(x,\Omega) = \frac{4\pi N_0(x)\alpha^2(x)}{\hbar N(x)} \int_{\Delta_g}^{\Omega-\Delta_g} [DoS_1(x,\varepsilon')DoS_1(x,\Omega-\varepsilon') + \mathrm{Im}F_1(x,\varepsilon')\mathrm{Im}F_1(x,\Omega-\varepsilon')]d\varepsilon',$$

(14)

where $\alpha^2$ is the square of the matrix element of the electron-phonon interaction and N is the ion number density of the material. In general $\alpha^2$ depends on energy. Average values can be found in ref. [19]. We recall that the position dependences for $N_0$, $\alpha^2$ and N come from the fact that the electrode material is not homogeneous across the junction. Therefore the value is different depending on the nature of the material at position x.

The Cooper pair breaking rate can be averaged over the position in the bilayer x and over the energy interval $\Delta\varepsilon_\alpha$ in the same way as it was done before, to yield the Cooper pair breaking rate $\Gamma_{PB}(\Omega)$ as a function of phonon energy averaged over the energy intervals $\Delta\varepsilon_\alpha$. Figure 7 shows the Cooper pair breaking rate as a function of phonon energy.

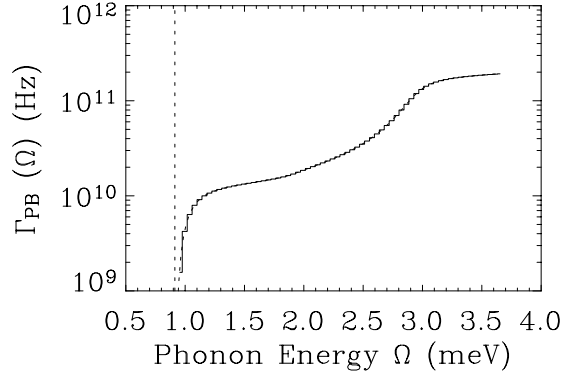

**Figure 7:**
Cooper pair breaking time, $\Gamma_{PB}(\Omega)$, as a function of phonon energy $\Omega$. The vertical dashed line represents the gap energy of the superconducting electrode.

### F. Recombination-mediated energy exchange in the quasiparticle system

Let us consider the following sequence of events: A quasiparticle from the energy interval $\Delta\varepsilon_\alpha$ recombines with a quasiparticle from the energy interval $\Delta\varepsilon_\beta$, thereby releasing a phonon of energy $\varepsilon_\alpha+\varepsilon_\beta$. This phonon then breaks a Cooper pair into two quasiparticles, one being released into the energy interval $\Delta\varepsilon_\gamma$ and the other into the interval $\Delta\varepsilon_{\alpha+\beta-\gamma}$, which corresponds to the energy interval around the energy $\varepsilon_\alpha+\varepsilon_\beta-\varepsilon_\gamma$.

This sequence annihilates two quasiparticles from the intervals $\Delta\varepsilon_\alpha$ and $\Delta\varepsilon_\beta$, and creates two quasiparticles in the intervals $\Delta\varepsilon_\gamma$ and $\Delta\varepsilon_{\alpha+\beta-\gamma}$. The rate for this sequence of events is given by[20]:

$$\Gamma_{ree}\left(x,\left(\Delta\varepsilon_\alpha,\Delta\varepsilon_\beta\right)\to\left(\Delta\varepsilon_\gamma,\Delta\varepsilon_{\alpha+\beta-\gamma}\right)\right)=\frac{\left(\varepsilon_\alpha+\varepsilon_\beta\right)^2\delta\varepsilon}{2N_0(x)\tau_0(x)(k_BT_C(x))^3V}\frac{4N_0(x)\pi\alpha^2(x)}{\hbar N(x)}\left[\frac{1}{\Gamma_{esc}+\Gamma_{PB}(x,\Delta\varepsilon_{\alpha+\beta})}\right]\cdot$$

$$\left[DoS_1(x,\Delta\varepsilon_\alpha)\cdot DoS_1(x,\Delta\varepsilon_\beta)+Im F_1(x,\Delta\varepsilon_\alpha)\cdot Im F_1(x,\Delta\varepsilon_\beta)\right]\cdot$$
$$\left[DoS_1(x,\Delta\varepsilon_\gamma)\cdot DoS_1(x,\Delta\varepsilon_{\alpha+\beta-\gamma})+Im F_1(x,\Delta\varepsilon_\gamma)\cdot Im F_1(x,\Delta\varepsilon_{\alpha+\beta-\gamma})\right]$$
(15)

where $\Gamma_{esc}$ is the phonon escape rate out of the film, V is the volume of the electrode and $DoS(x,\Delta\varepsilon)$ is the average density of states in the interval $\Delta\varepsilon$ and at the position x:

$$DoS(x,\Delta\varepsilon_\alpha)=\frac{1}{\delta\varepsilon}\int_{\Delta\varepsilon_\alpha}DoS(x,\varepsilon)d\varepsilon. \quad (16)$$

The same holds for $Im F(x,\Delta\varepsilon_\alpha)$.

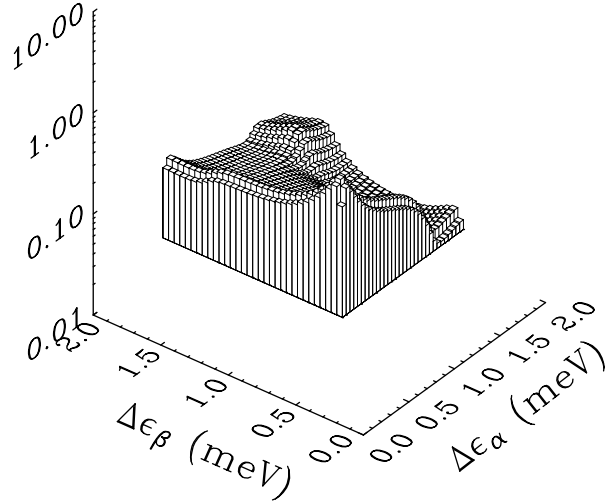

**Figure 8:**
Quasiparticle recombination with energy exchange rate, $\Gamma_{ree}\left(x,\left(\Delta\varepsilon_\alpha,\Delta\varepsilon_\beta\right)\to\left(\Delta\varepsilon_\gamma,\Delta\varepsilon_{\alpha+\beta-\gamma}\right)\right)$, as a function of the initial quasiparticle energies $\Delta\varepsilon_\alpha$ and $\Delta\varepsilon_\beta$, and for the final energy $\Delta\varepsilon_\gamma$ equal to the gap energy of the electrode.

This rate can then be averaged over the position x in the bilayer. Note that the rate depends on three independent indices $\alpha$, $\beta$ and $\gamma$, determining the energies of the two initial and one of the two final quasiparticles, the energy of the other final quasiparticle being fixed by the energy conservation law.

The phonon-mediated process of energy exchange in the electronic system of a superconductor, which we discussed above, may be treated exactly like an electron-electron collision process due to Coulomb interaction. We only need to disregard the short-lived intermediate pair-breaking phonon emitted in the initial collision of the pair of quasiparticles an thus consider the combined cross-section leading to (15) for scattering of the pair of colliding quasiparticles from their initial states into final states with the conservation of the total energy. Whilst the Coulomb interaction is important for establishing the equilibrium within the quasiparticle system at relatively large quasiparticle densities either near $T_C$ or at quasiparticle densities comparable to that of the condensate, the phonon mediated process considered above is by far the most important equilibration mechanism at small and moderate quasiparticle densities. Figure 8 shows the recombination mediated energy exchange rate in the quasiparticle system as a function of the initial quasiparticle energies $\Delta\varepsilon_\alpha$ and $\Delta\varepsilon_\beta$, and for the final energy $\Delta\varepsilon_\gamma$ equal to the gap energy of the electrode.

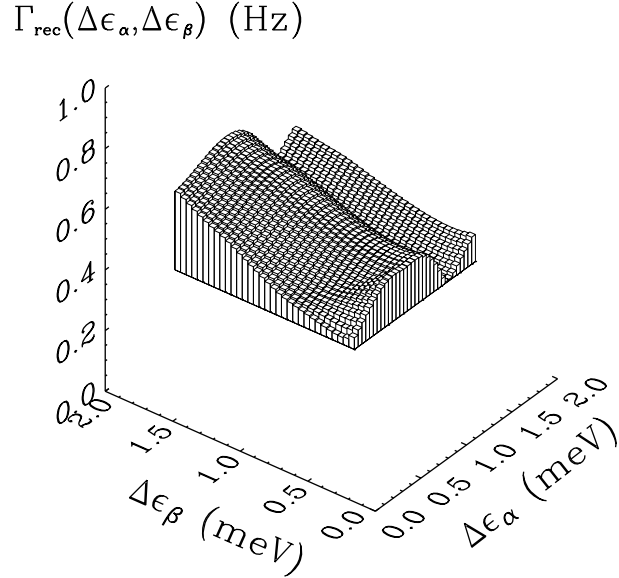

**Figure 9:**
Quasiparticle recombination with subsequent phonon loss rate, $\Gamma_{rec}(x,(\Delta\varepsilon_\alpha,\Delta\varepsilon_\beta))$, as a function of the initial quasiparticle energies $\Delta\varepsilon_\alpha$ and $\Delta\varepsilon_\beta$.

### G. Quasiparticle recombination

In the situation that the phonon released by a recombination process does not break another Cooper pair, but is lost into the substrate, the two quasiparticles are effectively lost from the system. This process is described by the following rate[20]:

$$\Gamma_{rec}(x,(\Delta\varepsilon_\alpha,\Delta\varepsilon_\beta)) = \frac{(\varepsilon_\alpha+\varepsilon_\beta)^2}{2N_0(x)\tau_0(x)(k_B T_C(x))^3 V}\left[\frac{\Gamma_{esc}}{\Gamma_{esc}+\Gamma_{PB}(x,\Delta\varepsilon_{\alpha+\beta})}\right]\left[\frac{DoS_1(x,\Delta\varepsilon_\beta)+\frac{\Delta_1(x)}{\varepsilon_\beta}\operatorname{Im}F_1(x,\Delta\varepsilon_\beta)}{DoS_1(x,\Delta\varepsilon_\beta)}\right]$$
(17)

Again, the rate can be averaged over the position in the bilayer x to yield the characteristic rate for quasiparticle recombination with phonon loss, depending on two indices α and β representing the energy of the two initial quasiparticles. For the practically important case when the recombination is bottlenecked due to fast phonon re-absorption, the recombination mediated energy exchange term is not. The probabilities of the two processes add up to unity. Thus for that case the energy exchange is a factor $\Gamma_{PB}/\Gamma_{esc} \gg 1$ faster than recombination.

Figure 9 shows the recombination rate as a function of the two quasiparticle energies.

### H. Quasiparticle trapping

Some regions in the superconducting film can have a local energy gap, which is lower than the energy gap of the superconductor surrounding it[21]. Possible reasons for a locally reduced gap are dislocations, single magnetic impurity atoms or their clusters giving discrete or continuous states inside the superconducting gap or small normal metal inclusions. A quasiparticle, which is close to such a region of reduced energy gap, can emit a phonon and scatter down to the lower energy gap region, where it is restrained from diffusing any further. Such a quasiparticle is trapped in the region of lower energy gap and is therefore effectively lost from the tunneling system. We can approximate the trapping rate of a quasiparticle in the energy interval $\Delta\varepsilon_\alpha$ as:

$$\Gamma_{trap}(\Delta\varepsilon_\alpha) = C_{trap}\Gamma_{emi}(\Delta\varepsilon_\alpha + d_{trap} \to \Delta_g),$$
(18)

where $d_{trap}$ is the trap depth, $C_{trap}$ is the trapping probability and $\Gamma_{emi}(\Delta\varepsilon_\alpha+d_{trap}\to\Delta_g)$ is the rate of quasiparticle scattering into the trap with emission of a phonon of energy $\Delta\varepsilon_\alpha-\Delta_g+d_{trap}$.

### I. Quasiparticle de-trapping

Phonon absorption can free a trapped electron out of the region of reduced energy gap and make it available to the tunnel system again. Similarly to the previous paragraph, we can write the detrapping rate by phonon absorption of a trapped quasiparticle into the energy interval $\Delta\varepsilon_\alpha$ as:

$$\Gamma_{deabs}(\Delta\varepsilon_\alpha) = \Gamma_{abs}(\Delta_g \to \Delta\varepsilon_\alpha + d_{trap}).$$
(19)

A second possibility for a quasiparticle to escape from the trap is via recombination with another quasiparticle from the electrode, having an energy $\Delta\varepsilon_\alpha$. Depending on the energy $\Delta\varepsilon_\alpha$ of this free quasiparticle the trapped quasiparticle is either completely lost from the electrode or freed from the trap. If the energy of the free quasiparticle is larger than $\Delta_g+d_{trap}$, the released phonon has enough energy to directly break a Cooper pair. The two quasiparticles created by this pair breaking process have an energy larger than the energy gap and can diffuse away from the trap:

$$\Gamma_{deree}(\Delta\varepsilon_\alpha, \Delta\varepsilon_\beta) = \Gamma_{ree}((\Delta_g, \Delta\varepsilon_\alpha) \rightarrow (\Delta\varepsilon_\beta, \Delta\varepsilon_{\Delta_g+\alpha-\beta})), \qquad \Delta\varepsilon_\alpha > \Delta_g + d_{trap}. \quad (20)$$

This second mechanism of trap ionization is analogous to shock ionization by thermal phonons. It is not linear and its rate depends on the quasiparticle density. At low temperatures and small non-equilibrium phonon densities this may become the only significant process.

### J. Quasiparticle loss

The quasiparticle loss rate $\Gamma_{loss}$ is modeled as being independent of the quasiparticle energy. It includes all other losses than the ones by trapping or recombination. An example of these direct losses is diffusion of the quasiparticles out of the junction area through the leads. In order to prevent quasiparticles from leaving the junction area the contacts to the base and top film are fabricated out of a higher $T_C$ material than the electrodes. In an ideal case these additional losses would be negligible compared to the trapping and recombination losses.

## III. Energy dependent balance equations

By regrouping all the terms calculated in the previous section, we can write the energy dependent balance equation of the QP number in the energy interval $\Delta\varepsilon_\alpha$ in the first electrode, $N_1(\Delta\varepsilon_\alpha)$:

$$\begin{aligned}
\frac{dN_1(\Delta\varepsilon_\alpha)}{dt} &= \Gamma_{tun,2}(\Delta\varepsilon_\alpha - eV_b \rightarrow \Delta\varepsilon_\alpha) \cdot N_2(\Delta\varepsilon_\alpha - eV_b) - \Gamma_{tun,1}(\Delta\varepsilon_\alpha \rightarrow \Delta\varepsilon_\alpha + eV_b) \cdot N_1(\Delta\varepsilon_\alpha) \\
&+ \Gamma_{can,2}(\Delta\varepsilon_\alpha + eV_b \rightarrow \Delta\varepsilon_\alpha) \cdot N_2(\Delta\varepsilon_\alpha + eV_b) - \Gamma_{can,1}(\Delta\varepsilon_\alpha \rightarrow \Delta\varepsilon_\alpha - eV_b) \cdot N_1(\Delta\varepsilon_\alpha) \\
&+ \sum_\beta \Gamma_{emi,1}(\Delta\varepsilon_\beta \rightarrow \Delta\varepsilon_\alpha) \cdot N_1(\Delta\varepsilon_\beta) - \sum_\beta \Gamma_{emi,1}(\Delta\varepsilon_\alpha \rightarrow \Delta\varepsilon_\beta) \cdot N_1(\Delta\varepsilon_\alpha) \\
&+ \sum_\beta \Gamma_{abs,1}(\Delta\varepsilon_\beta \rightarrow \Delta\varepsilon_\alpha) \cdot N_1(\Delta\varepsilon_\beta) - \sum_\beta \Gamma_{abs,1}(\Delta\varepsilon_\alpha \rightarrow \Delta\varepsilon_\beta) \cdot N_1(\Delta\varepsilon_\alpha) \\
&- \sum_\beta \sum_\gamma \Gamma_{ree,1}((\Delta\varepsilon_\alpha, \Delta\varepsilon_\beta) \rightarrow (\Delta\varepsilon_\gamma, \Delta\varepsilon_{\alpha+\beta-\gamma})) \cdot \left[ \frac{N_1(\Delta\varepsilon_\alpha) \cdot N_1(\Delta\varepsilon_\beta)}{DoS_1(\Delta\varepsilon_\alpha) \cdot DoS_1(\Delta\varepsilon_\beta)} - \frac{N_1(\Delta\varepsilon_\gamma) \cdot N_1(\Delta\varepsilon_{\alpha+\beta-\gamma})}{DoS_1(\Delta\varepsilon_\gamma) \cdot DoS_1(\Delta\varepsilon_{\alpha+\beta-\gamma})} \right] \\
&- \sum_\beta \Gamma_{rec,1}(\Delta\varepsilon_\alpha, \Delta\varepsilon_\beta) \cdot [N_1(\Delta\varepsilon_\alpha) \cdot N_1(\Delta\varepsilon_\beta) - N_{th,1}(\Delta\varepsilon_\alpha) \cdot N_{th,1}(\Delta\varepsilon_\beta)] \\
&- \Gamma_{trap,1}(\Delta\varepsilon_\alpha)[n_1^{traps} - N_1^t] \cdot N_1(\Delta\varepsilon_\alpha) + \Gamma_{det\,rap,1}(\Delta\varepsilon_\alpha) \cdot N_1^t + \sum_\beta \Gamma_{deree,1}(\Delta\varepsilon_\alpha, \Delta\varepsilon_\beta) \cdot \frac{N_1^t \cdot N_1(\Delta\varepsilon_\alpha)}{DoS_1(\Delta_g) \cdot DoS_1(\Delta\varepsilon_\alpha)} \\
&- \Gamma_{loss,1} \cdot N_1(\Delta\varepsilon_\alpha) \qquad (21)
\end{aligned}$$

where $N_{th,1}(\Delta\varepsilon_\alpha)$ is the number of thermal quasiparticles in the energy interval $\Delta\varepsilon_\alpha$ of electrode 1, $N_1^t$ is the number of trapped quasiparticles in electrode 1, $n_1^{traps}$ is the number of available traps in electrode 1 and $DoS(\Delta\varepsilon_\alpha)$ is the density of states in the energy interval $\Delta\varepsilon_\alpha$, averaged over the position in the bi-layer.

Of course we have one equation per energy interval $\Delta\varepsilon_\alpha$. This system of equations in electrode 1 has to be completed by the equation giving the variation of the number of trapped quasiparticles in electrode 1, which can be written as:

$$\frac{dN_1^t}{dt} = \sum_\alpha \left[ \Gamma_{trap,1}(\Delta\varepsilon_\alpha) \cdot [n_1^{traps} - N_1^t] \cdot N_1(\Delta\varepsilon_\alpha) - \Gamma_{det\,rap,1}(\Delta\varepsilon_\alpha) \cdot N_1^t \right]$$

$$- \sum_\alpha \sum_\beta \Gamma_{deree,1}(\Delta\varepsilon_\alpha, \Delta\varepsilon_\beta) \cdot \frac{N_1^t \cdot N_1(\Delta\varepsilon_\alpha)}{DoS_1(\Delta_g) \cdot DoS_1(\Delta\varepsilon_\alpha)} . \quad (22)$$

A similar set of equations can be written for electrode 2 by interchanging the indices 1 and 2. If there are $N_{en}$ energy intervals in one electrode, we end up with a system of $2N_{en}+2$ coupled, non-linear, first order differential equations, which has to be solved numerically. The numerical method used is a simple Euler iterations scheme, where the step size is varied from $10^{-11}$ seconds at the beginning of the pulse to approximately $10^{-8}$ sec at the end of the pulse. The variation of the step size is linked to the variation of the number of quasiparticles in the different energy intervals.

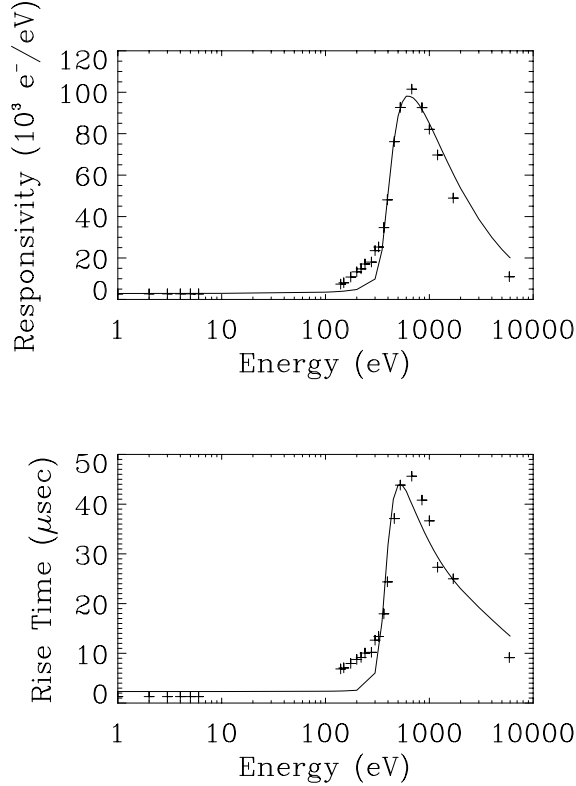

**Figure 10:**
Responsivity **(a)** and rise time **(b)** of the Nb-Al junction at a bias voltage of 180 µV as a function of incoming photon energy. The crosses represent the experimental data, whereas the solid line represents the calculated data with the parameters from table 2.

The initial conditions are found by remarking that the model is started after the second stage of the electronic down-conversion process[22], as defined in ref. [22]. At this point phonons created by relaxation of the quasiparticles cannot break Cooper pairs anymore. The number of quasiparticles $Q_0$ created by the first two stages of the down-conversion process was calculated to be equal to[22]:

$$Q_0 = \frac{E}{1.7\Delta_g} \quad , \qquad (23)$$

where E is the photon energy.

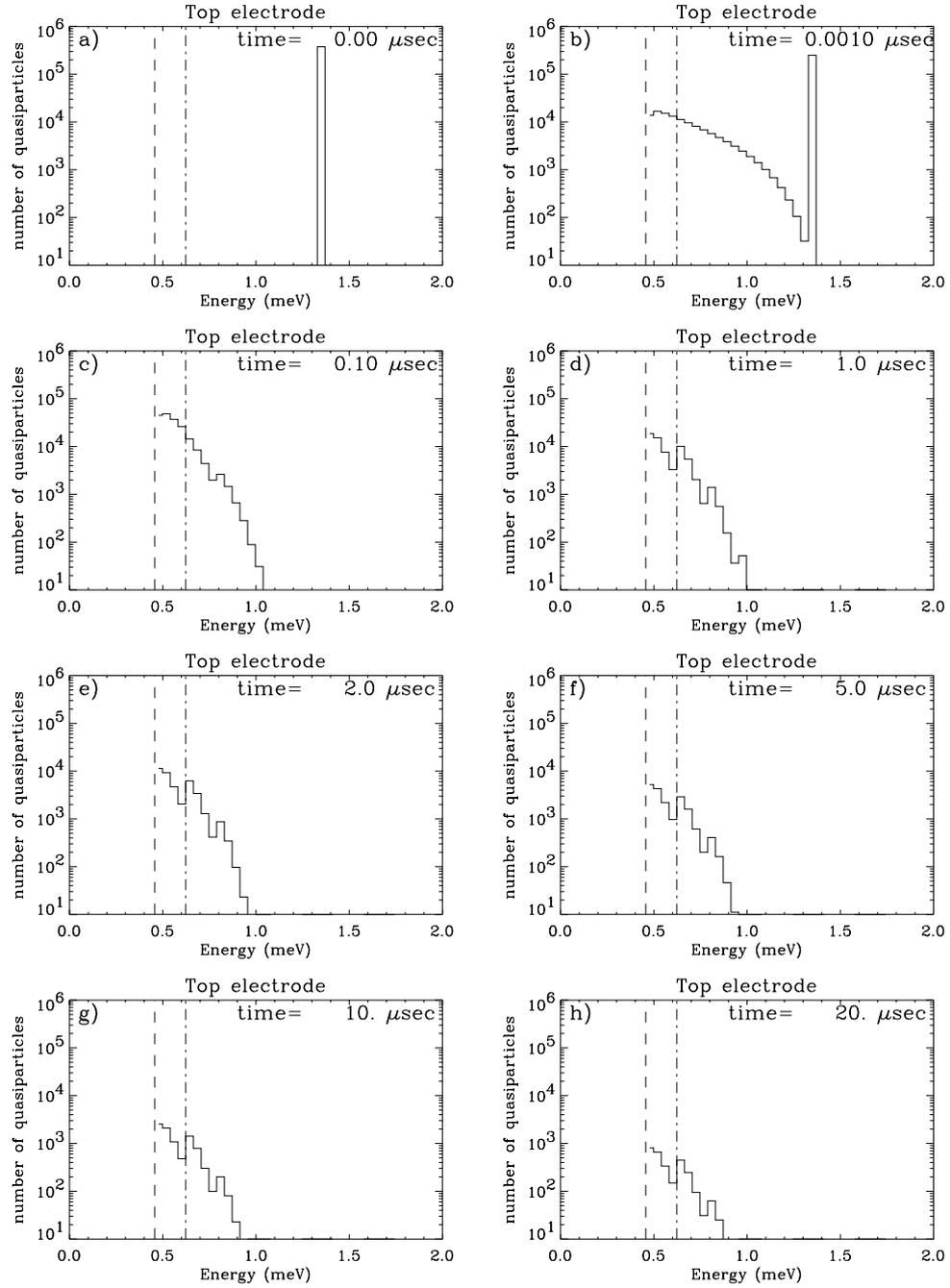

**Figure 11:**
Quasiparticle energy distribution for eight different instants after absorption of a 300 eV photon in the top electrode of the junction. In every graph the dashed vertical line represents the energy gap of the superconducting electrode of the junction, whereas the dashed-dotted vertical line represents the bias energy $eV_b$ above the energy gap.

Therefore, at t=0, this number of quasiparticles is put in the highest energy interval below $3\Delta_g$ of the absorbing electrode. The exact energy distribution at t=0 is of no importance, as the excess quasiparticles very quickly (~0.1 μsec) find a "quasi-equilibrium" distribution via tunneling and relaxation, totally independent of the initial distribution of the quasiparticles (see next section).

## IV.  Results of simulations

In order to illustrate the new model we describe in the following simulations made for a symmetrical Superconducting Tunnel Junction with 20 nm side length, where each electrode consists of a 100 nm thick Nb layer and a 120 nm thick Al layer. This junction was tested extensively at 350 mK in a portable $^3$He sorption cryostat at the BESSY synchrotron radiation facility from 30 to 2000 eV, in the optical regime from 1 to 5 eV and with an $^{55}$Fe radiation source emitting 6keV radiation. The variation of the responsivity (charge output per unit of incoming photon energy) and the pulse decay time, which is equal to the rise time of the output of the charge sensitive pre-amplifier, were measured as a function of incoming photon energy at a bias voltage of 180 μeV. The experimental data are shown in figure 10.

The model has five unknown fitting parameters, which are: the energy independent quasiparticle loss rate $\Gamma_{loss}$, the energy independent phonon escape rate from the electrode $\Gamma_{esc}$, the trapping probability $C_{trap}$, the number of available traps in the electrode $n^{traps}$ and the trap depth $d_{trap}$. All of the fitting parameters have a profound physical interpretation and all of them, except for $C_{trap}$ and $d_{trap}$, have a single, very specific region of influence on the curve in figure 10, which allows them to be determined with certainty:

   i-  The quasiparticle loss time $\tau_{loss}=1/\Gamma_{loss}$ reflects the rise-time of the signal, when all other loss channels, like losses by trapping and losses by recombination, are negligible. Therefore, it determines the height of the maximum of the curve, when all the trapping states are saturated and filled with quasiparticles and quasiparticle recombination has not yet set in. In case the losses by trapping and recombination are dominant over the whole energy range this fitting parameter can be securely neglected.
   ii- The phonon escape rate out of the electrode $\Gamma_{esc}$ determines the losses by quasiparticle recombination, which only become dominant at high quasiparticle densities in the electrodes of the junction, and therefore only influences the steepness of the negative slope of the curve in the high photon energy domain.
   iii- The number of available traps in the electrode $n^{traps}$ determines the onset of the positive slope part of the curve, when the quasiparticle traps start to be saturated with quasiparticles. The more available traps there are, the more the onset of the positive slope is shifted towards high photon energies, because it takes more quasiparticles to fill the available traps.

- iv- The trapping probability $C_{trap}$ influences the speed at which a quasiparticle is trapped. This parameter mainly determines the decay time of the pulse in the low energy part of the curve, by setting the speed at which a small number of quasiparticles is lost into a large amount of traps. This region of the curve is linear, as the traps are far from being saturated.
- v- The trap depth $d_{trap}$ has the same influence to the trapping speed as the trapping probability $C_{trap}$. As these two parameters both act in the same direction, it is fairly difficult to determine these two characteristics in an absolute manner just with the data of figure 10. Nevertheless, the trap depth can be derived from other experiments, like from responsivity and rise-time measurements as a function of temperature[23].

The preceding five parameters were varied in order to find a fit to the experimental data of figure 10. The calculated curves are also shown in the figure. Table II shows the set of parameters associated to the 20 μm Nb-Al junction, which were determined to fit the results of our model to the experimental data of figure 10.

**Table II. Fitting parameters of the model.**

| Symbol | Name | Unit | Value |
|---|---|---|---|
| $\tau_{loss}=1/\Gamma_{loss}$ | Quasiparticle loss time | μsec | 100 |
| $\Gamma_{esc}$ | Phonon escape rate | Hz | $5\,10^9$ |
| $C_{trap}$ | Trapping probability | / | 0.22 |
| $n^{traps}$ | Number of traps in electrode | / | 185 000 |
| $d_{trap}$ | Trap depth | μeV | 240 |

Let us now discuss the main scope of the paper, the quasiparticle energy distribution and its variation with time.

### A. Convergence to a "quasi-equilibrium" distribution

After 0.1 to 0.5 μsec the quasiparticle distribution converges from its initial state to a "quasi-equilibrium" distribution. This distribution is called to be in "quasi-equilibrium" in the sense that the normalized energy distribution of the quasiparticles stays constant, whereas the total number of quasiparticles in the electrodes diminishes, because of the different quasiparticle loss channels. This is illustrated in Fig. 11, where the quasiparticle energy distribution is shown at different instants of time in the absorbing electrode after the absorption of a 300 eV photon.

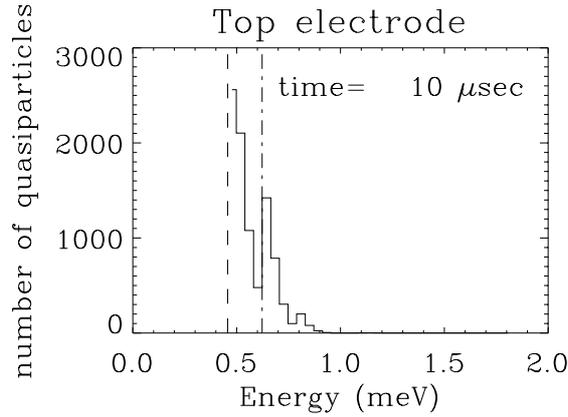

**Figure 12:**
Quasiparticle energy distribution on a linear scale, ten microseconds after the absorption of a 300 eV photon in the top electrode of the junction. The dashed vertical line represents the energy gap of the superconducting electrode of the junction, whereas the dashed-dotted vertical line represents the bias energy $eV_b$ above the energy gap. The step-like structure of the quasiparticle distribution can be clearly identified.

At the moment of photon absorption (t=0) all the quasiparticles are in an elevated energy level of the absorbing electrode. An exact knowledge of the initial conditions is not necessary, as, through a series of phonon scattering and tunnel events, the quasiparticle population converges within a fraction of a microsecond towards a stable configuration. One can see on the figure that for times t>1 μsec, the shape of the distribution does not vary anymore. Only the absolute number of quasiparticles decreases with time because of the different quasiparticle loss channels. This "quasi-equilibrium distribution" shows a step-like structure, caused by the discrete energy gain (loss) of $eV_b$ during a tunnel (cancellation-tunnel) process. In figure 12, which shows the quasiparticle energy distribution at t=10 μsec on a linear scale, this step-like structure can be seen more clearly. The energy difference between two consecutive steps is $eV_b$. This structure is similar to the results of Kozorezov et al., who already observed the step-like structure of the quasiparticle energy distribution for BCS-like junctions in thermal equilibrium[23]. In their case this particular structure was responsible for current steps in the IV-curves of the junctions. Segall et al.[24], who made a similar, but less complete code for the special case of BCS-like junctions, do not mention the step-like nature of the quasiparticle distribution. They compare the quasiparticle energy distribution at one microsecond to a thermal distribution with an effective temperature $T_{eff}$, higher than the bath temperature. The junctions for which they make their computations have relatively long tunnel times of the order of 2 μsec, and they only give the quasiparticle distribution at a time t=1μsec, which is smaller than the tunnel time. This is why the steps are not yet visible in their graphs, but should build up for times larger than the tunnel time. Therefore, their results are not in contradiction to our findings.

## B. Tunnel Current

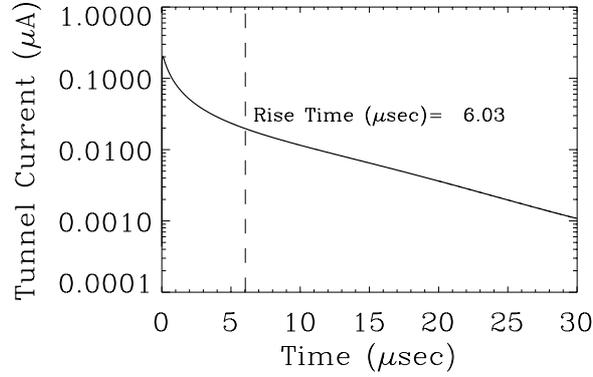

**Figure 13:**
Tunnel current pulse on a logarithmic scale as a function of the time after the absorption of a 300 eV photon in the top electrode of the Nb-Al junction. The applied bias voltage is 180 µeV. The vertical dashed line represents the decay time of the pulse.

The tunnel current is given by the sum of the tunnel terms minus the sum of the cancellation terms. After reaching a maximum within a fraction of a microsecond, the tunnel current decays mainly exponentially. Deviations from the simple exponential decay are observed in case of high losses by recombination and in case of trap saturation during the pulse. Figure 13 shows the tunnel current pulse from the Nb-Al junction under discussion after the absorption of a 300 eV photon on a logarithmic scale. One can clearly identify that the pulse does not decay purely exponentially at the beginning of the pulse because of fast quasiparticle trapping and fast quasiparticle recombination. Figure 14 shows the integrated current pulse, representing the charge output of the junction. In order to determine the charge output accurately, the pulse has to be calculated up to several times the decay time. In order to shorten the calculations, the last part of the current pulse is fitted with an exponentially decaying curve of the form:

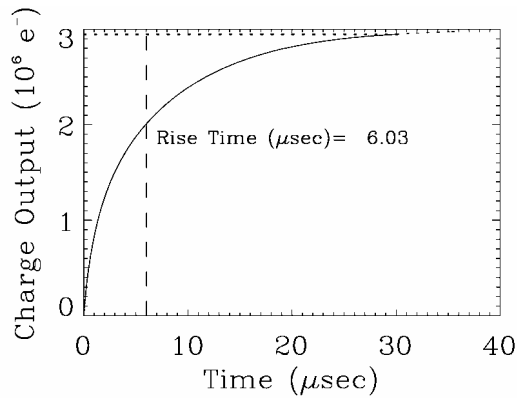

**Figure 14:**
Integrated current pulse as a function of time after the absorption of a 300 eV photon in the top electrode of the Nb-Al junction. The applied bias voltage is 180 µeV. The vertical dashed line represents the rise time of the integrated pulse and the horizontal dashed line represents the total charge output of the junction.

$$I(t) = I_m \exp(-t/\tau_D). \tag{24}$$

The calculation can then be stopped after a time slightly larger than the decay time of the pulse. The total charge output and decay time are then determined from the combined calculated and fitted curves.

### C. Average values and tunnel over cancellation ratio

Knowing the quasi-particle energy distribution during a current pulse, we can now determine the average quasi-particle energy and the average characteristic tunnel times according to:

$$\langle \Gamma \rangle = \frac{\sum_i N(\Delta\varepsilon_i) \Gamma(\Delta\varepsilon_i)}{\sum_i N(\Delta\varepsilon_i)}, \tag{25}$$

where $\Gamma(\Delta\varepsilon_i)$ is a characteristic parameter in the energy interval $\Delta\varepsilon_i$ and $N(\Delta\varepsilon_i)$ is the number of quasi-particles in the interval $\Delta\varepsilon_i$. From these values the tunnel over cancellation ratio $\sigma = \frac{\langle \Gamma_{tun} \rangle}{\langle \Gamma_{can} \rangle}$ can be derived, which characterizes the fraction of charge output lost due to the cancellation tunnel events. The knowledge of this ratio, which cannot be determined experimentally and is important in order to derive for example the cancellation noise (Ref. 10), allows us to determine the mean number of tunnel and cancellation tunnel events per quasiparticle from the charge output (Q) of the junction:

$$\langle n_{tun} \rangle = \frac{\sigma}{\sigma - 1} \frac{Q}{Q_0}, \tag{26}$$

$$\langle n_{can} \rangle = \frac{1}{\sigma - 1} \frac{Q}{Q_0}, \tag{27}$$

$$\langle n \rangle = \langle n_{tun} \rangle + \langle n_{can} \rangle = \frac{\sigma + 1}{\sigma - 1} \frac{Q}{Q_0}, \tag{28}$$

where $Q_0$ is the initial number of quasi-particles created in the electrode.

Figure 15 shows how the average quasiparticle energy, the average tunnel time, the average cancellation tunnel time and the tunnel over cancellation ratio vary during the current pulse. Clearly, as the quasiparticle energy distribution converges towards the "quasi-equilibrium" distribution, the four characteristics also converge. Note that, even in the "quasi-equilibrium" distribution, the quasiparticles have an average energy of 570 µeV, which is considerably higher than the gap energy of the electrode (450 µeV). This feature was already predicted qualitatively by A. Poelaert et al. by taking advantage of a non energy dependent kinetic model (Ref. 12). We can thus talk of a quasiparticle heating effect, caused by the energy gain due to tunneling. Of course the exact value of the average quasiparticle energy will depend strongly on the tunneling time, the bias voltage and the energy gap of the superconductors forming the electrodes.

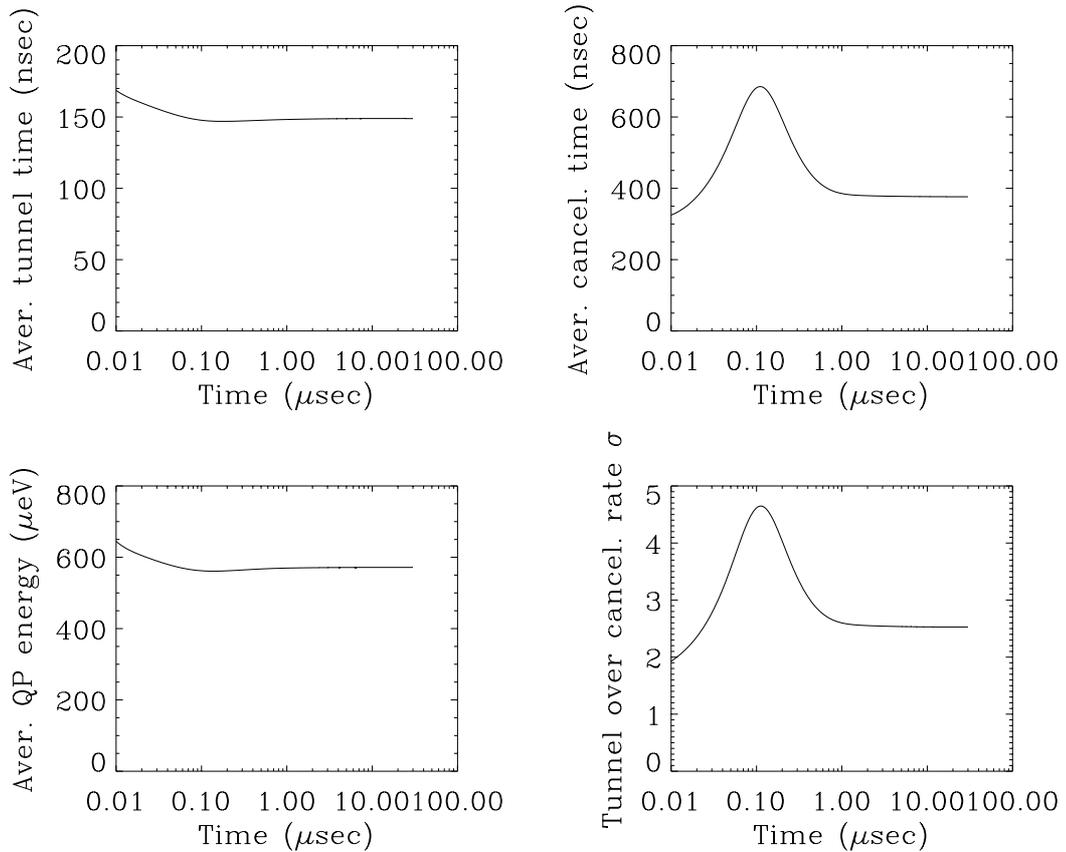

**Figure 15:**
Average quasiparticle energy **(a)**, average tunnel time **(b)**, average cancellation tunnel time **(c)** and tunnel over cancellation ratio **(d)** as a function of time after the absorption of a 300 eV photon in the top electrode of the Nb-Al junction. The applied bias voltage is 180 μeV. All four variables converge after a microsecond. The time scale is exponential in order to be able to see the very rapid convergence towards the "quasi-equilibrium" values.

## V. Conclusions

We have presented an energy dependent kinetic model of photon absorption in the most general type of superconducting tunnel junction, the inhomogeneous device. Based on the results of proximity effect theory, we presented expressions for all major quasiparticle processes occurring in the junctions as a function of initial and final quasiparticle energies and using a system of kinetics equations calculated the quasiparticle number variations as a function of energy and time.

We then applied the model to a symmetric Nb-Al superconducting tunnel junction with 100 nm of Nb and 120 nm of Al. By fitting the calculated values to experimental data, showing the responsivity and rise time as a function of photon energy, we were able to determine five characteristics of the junction, which are the quasiparticle trap number and depth, the quasiparticle trapping probability and the phonon and quasiparticle loss times. The model is very consistent in itself as the model has only five unknown fitting parameters, which all have a profound physical interpretation and can be determined with certainty from the different experimental curves.

The model shows that the quasiparticle energy distribution converges rapidly to a "quasi-equilibrium" distribution, in the sense that only the absolute number of quasiparticles varies in the energy intervals, whereas the distribution over the energy intervals remains constant. This "quasi-equilibrium" distribution has a step-like structure, where the energy difference between subsequent steps is equal to the bias energy $eV_b$. The model also shows that the mean quasiparticle energy lies well above the gap energy of the electrodes. The preceding results will be helpful for the understanding of the behavior of superconducting tunnel junctions used as photon detectors. Especially in the new generation of junctions fabricated out of low $T_C$ superconductors, the down-scattering time is large and as a consequence the quasiparticles will be spread over a wide energy domain. For a correct description and understanding of these low $T_C$ junctions the usual approach based on a model by Rothwarf and Taylor is not sufficient anymore and the presented model is essential. The description of data from these low $T_C$ junctions will be treated in future publications.